\documentclass[conference]{IEEEtran}
\IEEEoverridecommandlockouts
\usepackage{cite}
\usepackage{amsmath,amssymb,amsfonts}
\usepackage{algorithmic}
\usepackage{graphicx}
\usepackage{textcomp}
\usepackage{xcolor}
\usepackage{adjustbox}
\usepackage{lipsum}
\ifCLASSOPTIONcompsoc
    \usepackage[caption=false, font=normalsize, labelfont=sf, textfont=sf]{subfig}
\else
\usepackage[caption=false, font=footnotesize]{subfig}
\fi

\def\BibTeX{{\rm B\kern-.05em{\sc i\kern-.025em b}\kern-.08em
    T\kern-.1667em\lower.7ex\hbox{E}\kern-.125emX}}

\renewcommand\IEEEkeywordsname{Keywords}

\begin{document}

\title{Effect of Dispersion and Different Carrier Transitions on Absorption Characteristics of GaAs
}

\author{\IEEEauthorblockN{1\textsuperscript{st} Subarna Alam}
\IEEEauthorblockA{\textit{Department of EEE} \\
\textit{Bangladesh University of Engineering and Technology}\\
Dhaka, Bangladesh \\
1606082@eee.buet.ac.bd}
\and
\IEEEauthorblockN{2\textsuperscript{nd} Mohammad Jahangir Alam}
\IEEEauthorblockA{\textit{Department of EEE} \\
\textit{Bangladesh University of Engineering and Technology}\\
Dhaka, Bangladesh \\
mjalam@eee.buet.ac.bd}
}

\maketitle

\begin{abstract}
One of the basic optoelectronic characteristics is the absorption of any optoelectronic device or material. We present the characteristic of pure GaAs and the deviation of ideal spectra. Due to crystal defects, vacancies, impurities, etc., energy states exist in the semiconductor other than the allowed bands (conduction and valence bands). In this research, the effect of transitions between these states on absorption spectra and how the absorption coefficients change with the density and location of those states are analyzed numerically. Among the cases of 0.07 $eV$, 0.12 $eV$, and 0.17 $eV$ with the density of 2\%, 5\%, 10\%, 20\%, and 30\%, the minimum deviation occurs for the states located 0.07 
 $eV$ below the conduction band and when the energy states have the minimum density of 2\% and the deviation increases for the cases of 0.12 $eV$ and 0.17 $eV$ with higher densities. 

\end{abstract}


\section{Introduction}
Over the past decades, researchers have been focusing more on studying compound semiconductors, especially GaAs and other III-V materials, as the demand is increasing rapidly \cite{zeng2023recent}. GaAs provides many superior properties to those of Si. Because of being direct band gap material, GaAs can absorb and emit light more efficiently. Moreover, this material is less susceptible to radiation damage because of its wider bandgap \cite{anspaugh1984radiation}. Many annealing techniques have been applied to minimize the damage caused by electron and proton radiation \cite{loo1990radiation}. Another of its many uses is in producing laser diodes \cite{hall1962tj}.\\

\begin{figure}[ht] 
    \centering
  \subfloat[\label{eb1}]{%
        \includegraphics[width=0.45\linewidth]{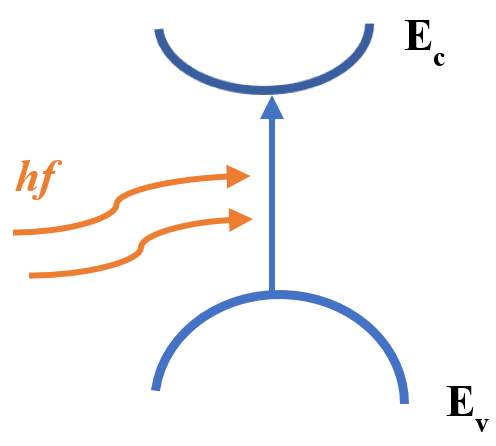}}
    \hfill
  \subfloat[\label{eb2}]{%
        \includegraphics[width=0.51\linewidth] {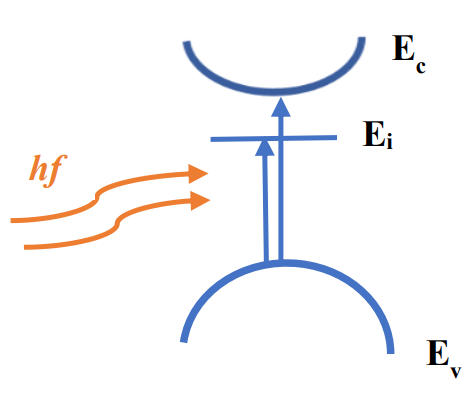}}
  \caption{Band diagram and absorption in (a) pure direct band gap material, (b) direct band gap material with energy states exist at $E_{i}$ }
  \label{energyband} 
\end{figure}

In any ideal semiconductor, there is no energy state in the forbidden gap. Fig. \ref{eb1} illustrates the energy band diagram and absorption for a pure direct band gap material. But in practical semiconductors, there are impurities, defects, vacancies, etc. As a result, there exist forbidden energy states. Fig. \ref{eb2} shows the same diagram if energy states exist at level $E_{i}$. In this case, transitions between $E_{v}$ and $E_{i}$ will also exist. Any ideal semiconductor characteristic gets deviated because of those transitions.\\

In this paper, the absorption spectra of GaAs and different factors that alter the ideal spectra have been studied. The absorption coefficients for the case of Fig. \ref{eb1} and Fig. \ref{eb2} have been evaluated.

\section{Methods and Experimental Analysis}
There are upward (absorption) and downward (emission) transmissions between energy bands in optical devices or semiconductors. Absorption allows electrons to transit from a lower energy state to a higher state. In direct band-to-band absorption, the electrons will transit from the valence band to the conduction band. Direct band-to-band absorption in intrinsic GaAs is related by Equation \eqref{eq1} \cite{bhattacharya1997semiconductor}.
\begin{equation}
\alpha  \propto \dfrac{\sqrt{hf - E_{g}}}{nhf}
\label{eq1}
\end{equation}


Where $\alpha$ is the absorption coefficient of GaAs, $f$ is the incident frequency, $E_{g}$ is the band gap of GaAs which is 1.42$eV$ in this analysis, and $n$ is the refractive index.
The refractive index is also an energy-dependent parameter \cite{marple1964refractive}. The absorption spectra are analyzed again, considering this dispersion.
Finally, the effects of forbidden states as illustrated in Fig. \ref{eb2} on absorption are analyzed as a function of forbidden state density at different levels.

\begin{figure}[ht]
  \centering
  \includegraphics[width=0.7\linewidth]{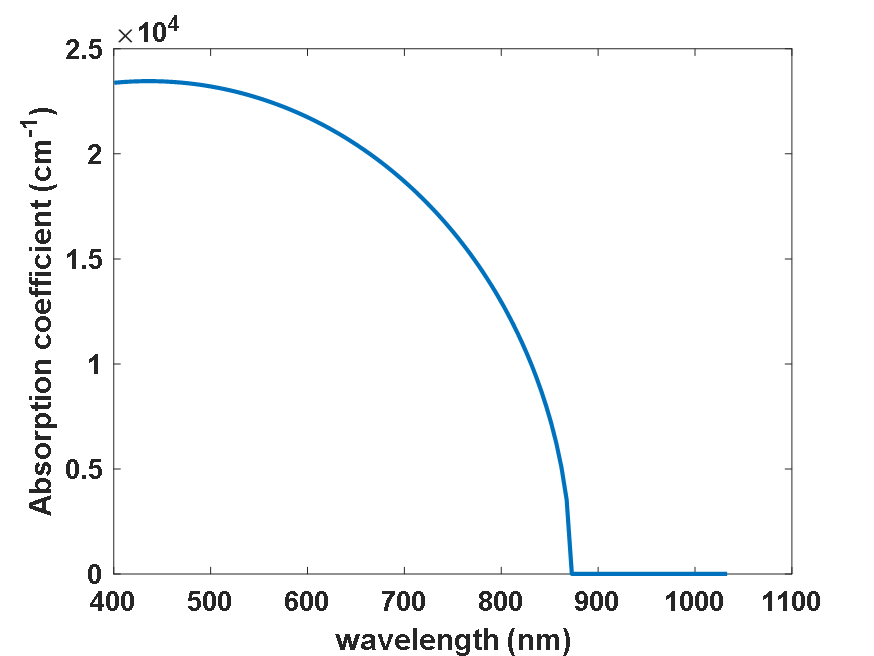}
  \caption{Absorption coefficient vs wavelength considering no dispersion or keeping refractive index constant}
  \label{fig1}
\end{figure}

\begin{figure}[ht]
    \centering
  \subfloat[\label{2a}]{%
        \includegraphics[width=0.5\linewidth]{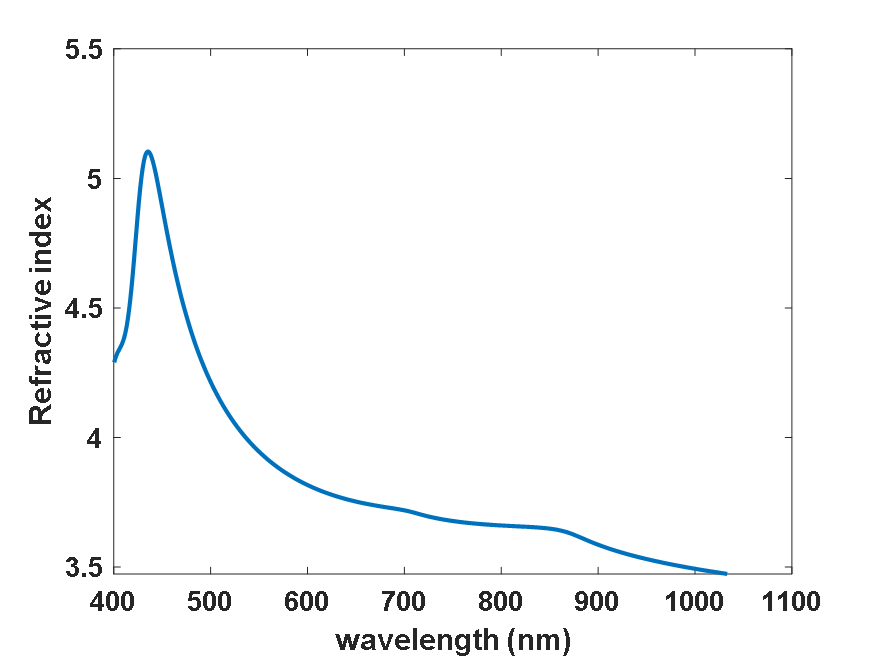}}
    \hfill
  \subfloat[\label{2b}]{%
        \includegraphics[width=0.5\linewidth] {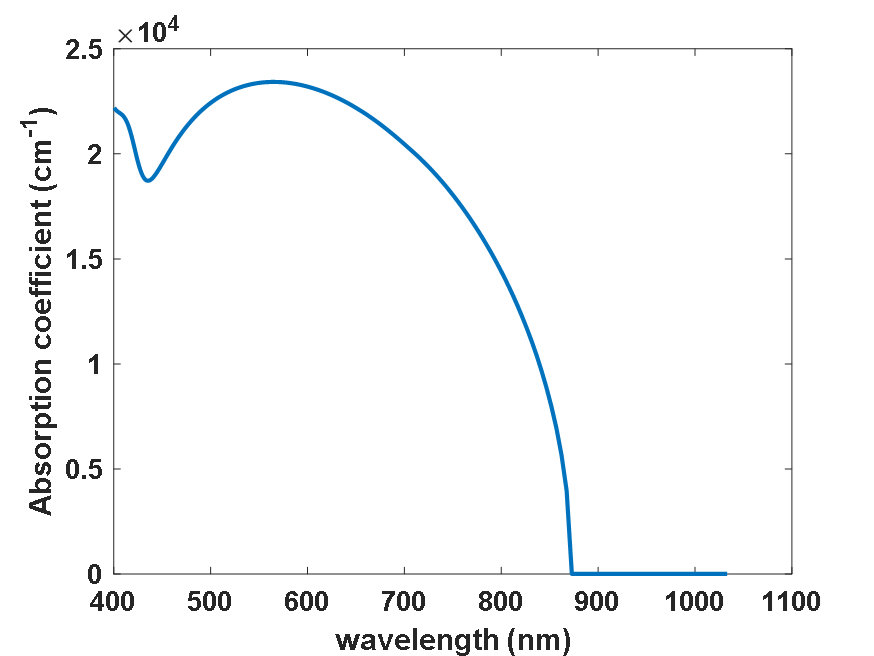}}
  \caption{(a) Wavelength dependency of Refractive index of GaAs \cite{marple1964refractive, rakic1996modeling} and (b) Absorption co-efficient vs wavelength considering dispersion (variable refractive index)}
  \label{fig2} 
\end{figure}

\section{Result and Discussion}
Fig. \ref{fig1} shows the absorption spectrum of intrinsic GaAs with constant refractive index, i.e., no dispersion. The refractive index is a nonlinear function of wavelength as per Fig. \ref{2a}, and so is the absorption coefficient. Fig. \ref{2b} shows the absorption characteristics of pure GaAs where the refractive index also changes with the wavelength. No energy below 1.42 $eV$ (bandgap of GaAs) gets absorbed in pure GaAs. So, no coefficient is achieved beyond 872.45 nm. 

\begin{figure}[ht] 
    \centering
  \subfloat[\label{3a}]{%
        \includegraphics[width=0.6\linewidth]{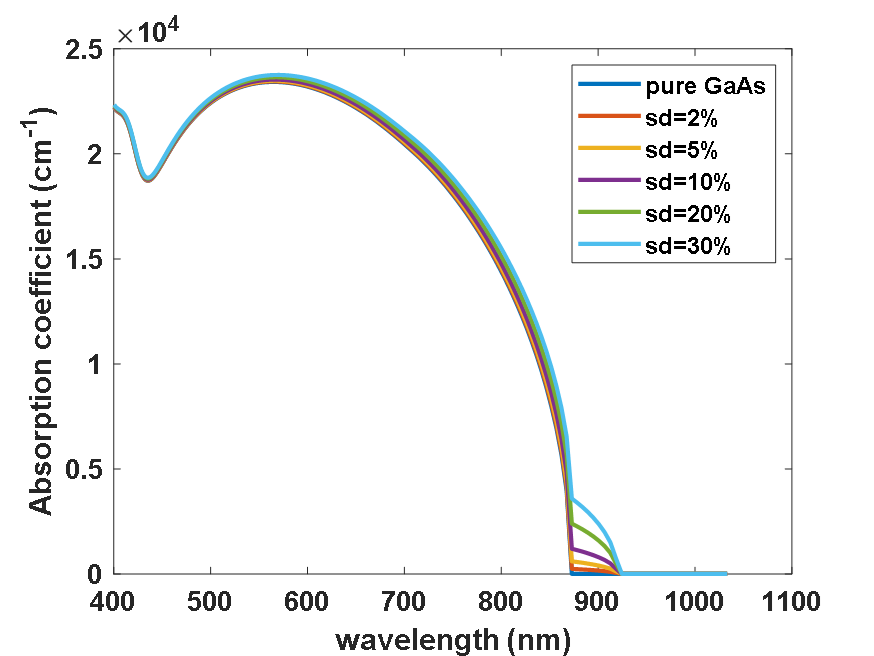}}
    \\
  \subfloat[\label{3b}]{%
        \includegraphics[width=0.49\linewidth]{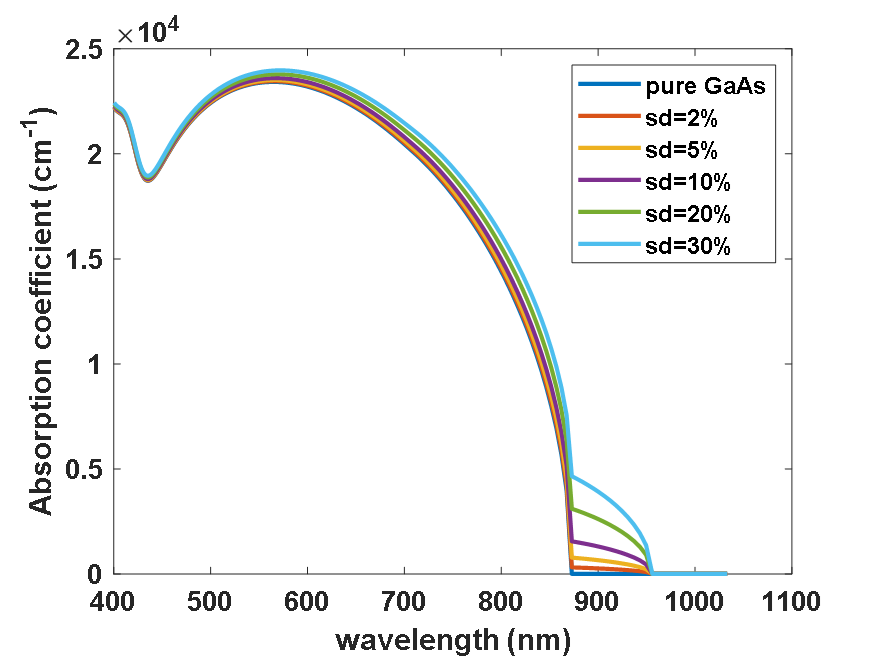}}
    \hfill
  \subfloat[\label{3c}]{%
        \includegraphics[width=0.49\linewidth]{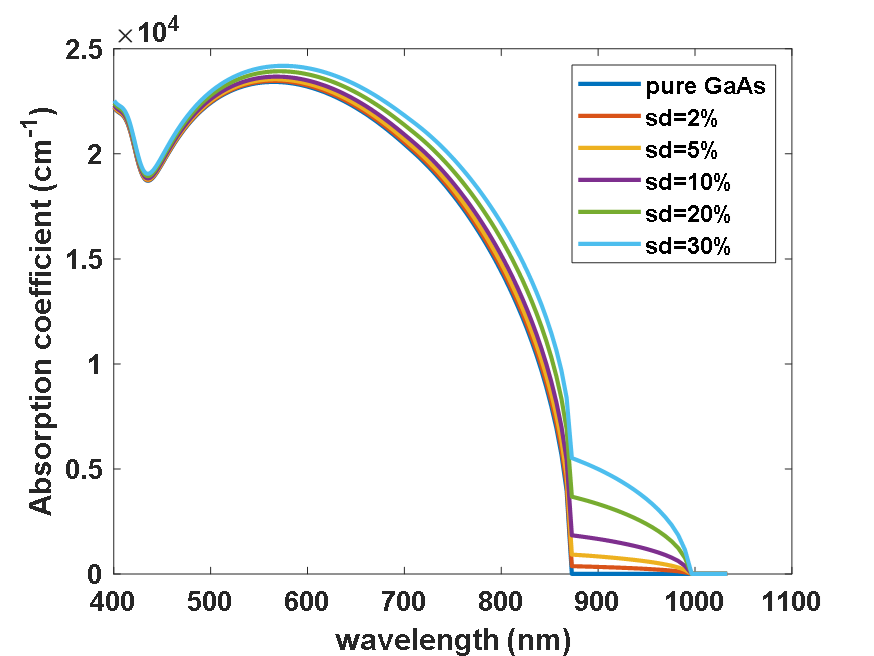}}
  \caption{Absorption coefficients vs wavelength considering the presence of energy states (a) 0.07$eV$, (b) 0.12$eV$, and (c) 0.17$eV$, below the conduction band where ‘sd’ represents the density of those states}
  \label{fig3} 
\end{figure}

Fig. \ref{fig3} shows the characteristics when energy states exist at different levels other than the conduction band and the valence band. Curves get deviated in the presence of those states. Lower energies than 1.42$eV$ are also absorbed in the presence of those states. Fig. \ref{3a} indicates that there is absorption up to the wavelength of ~925 nm. The same wavelengths for the cases of Fig. \ref{3b} and Fig. \ref{3c} are ~956 nm and ~993 nm, respectively. The following two important observations can be stated from Fig. \ref{fig3}:
\begin{itemize}
    \item The wider the gap between the conduction band and impurity energy states, the more the absorption characteristics will get deviated from the ideal ones.
    \item The higher the density of those energy states, the more the deviation will occur.
\end{itemize}

\section{Conclusion}
In this study, absorption characteristics and the various factors that affect the characteristics have been analyzed. In device processing, 100\% of purity cannot be achieved. So, the ideal characteristics cannot be achieved either. But the characteristic of available practical materials or devices can be compared with that of pure materials or devices. Absorption spectra studied above can be an indicator to measure the energy level of impurities, vacancies, and defects and the density of those states.

\section*{Acknowledgment}
The authors would like to thank Dr. Muhammad Anisuzzaman Talukder, Professor, Department of EEE, BUET, for his instructions in the optoelectronics lab.

\bibliographystyle{IEEEtran}
\bibliography{conference_101719.bib}



\end{document}